\documentclass[10pt,conference]{IEEEtran}
\IEEEoverridecommandlockouts
\usepackage{cite}
\usepackage{amsmath,amssymb,amsfonts}
\usepackage{graphicx}
\usepackage{textcomp}
\usepackage{xcolor}
\def\BibTeX{{\rm B\kern-.05em{\sc i\kern-.025em b}\kern-.08em
    T\kern-.1667em\lower.7ex\hbox{E}\kern-.125emX}}

\usepackage[utf8]{inputenc}
\usepackage{color}
\usepackage[T1]{fontenc}
\usepackage[english]{babel}

\usepackage{url}

\usepackage{listings}

\usepackage{authblk}

\usepackage{caption}
\usepackage{subcaption}
\usepackage[official]{eurosym}

\usepackage{cleveref}
\crefname{section}{§}{§§}
\Crefname{section}{§}{§§}

\usepackage{setspace}
\usepackage{float}
\usepackage{cuted}
\usepackage{flushend}

\begin{document}

\title{Distributed Resource Allocation and Application Deployment in Mesh Edge Networks}

%\author{Antoine BERNARD, Sandoche BALAKRICHENAN, Michel MAROT}
\author[1,3,4]{Antoine Bernard}
\author[1,2,3]{Antoine Legrain}
\author[4]{Maroua Ben Attia}
\author[4]{Abdo Shabah}

%\affil[1]{CIRRELT - Interuniversity Research Center on Enterprise Networks, Logistics}
%\affil[ ]{and Transportation, Montréal, Québec, Canada}
%\affil[2]{GERAD - Group for Research in Decision Analysis, Montréal, Québec, Canada}
%\affil[3]{Polytechnique Montréal, Montréal, Québec, Canada}
\affil[1]{CIRRELT / $^2$GERAD / $ ^3$Polytechnique Montréal, Montréal, Québec, Canada}
\affil[4]{Humanitas Solutions, Montréal, Québec, Canada}
\affil[ ]{\textit{antoine.bernard@polymtl.ca, antoine.legrain@polymtl.ca}}
\affil[ ]{\textit{maroua@humanitas.io, abdo@humanitas.io}}

\maketitle

\begin{abstract}
Virtual Network Embedding (VNE) approaches typically assume static or slowly-changing network topologies, but emerging applications require deployment in mobile environments where traditional methods become insufficient. This work extends VNE to constrained mesh networks of mobile edge devices, addressing the unique challenges of rapid topology changes and limited resources. We develop models incorporating device capabilities, connectivity, mobility and energy constraints to evaluate optimal deployment strategies for mobile edge environments. Our approach handles the dynamic nature of mobile networks through three allocation strategies: an integer linear program for optimal allocation, a greedy heuristic for immediate deployment, and a multi-objective genetic algorithm for balanced optimization. Our initial evaluation analyzes application acceptance rates, resource utilization, and latency performance under resource limitations. Results demonstrate improvements over traditional approaches, providing a foundation for VNE deployment in highly mobile environments.
\end{abstract}

\textbf{Keywords - Mobile edge computing, Virtual network embedding, Drone networks, Resource allocation}

\section{Introduction}

%% Reduced version
Virtual Network Embedding (VNE) has been extensively studied for cloud and datacenter environments, but emerging applications increasingly require deployment in highly mobile environments where traditional approaches become insufficient. Mobile edge computing scenarios, including drone networks, vehicular computing, and emergency response systems, present unique challenges that existing VNE methods cannot adequately address.

This work focuses on optimizing application deployment and resource management in highly dynamic, constrained environments, particularly drone-based mesh edge networks. Applications in such environments include emergency response coordination, mobile surveillance systems, and dynamic edge computing services where infrastructure may be unavailable.

We extend VNE beyond datacenters by integrating device capabilities, connectivity, mobility, and application requirements into a unified framework. We evaluate three allocation strategies: an integer linear program for optimal allocation, a greedy heuristic for immediate deployment, and a multi-objective genetic algorithm for balanced optimization.

We benchmark our system with fixed arrival and departure rates, establishing a baseline for understanding how conventional VNE approaches perform when applied to highly mobile environments.
Our primary contributions are:
\begin{itemize}
    \item An integer linear program formulation for optimal resource allocation in mobile distributed networks.
    \item Comparative evaluation of three allocation strategies under realistic positioning and bandwidth constraints.
    \item A comprehensive simulation framework for evaluating VNE in highly mobile environments as a foundation for real-world deployment.
\end{itemize}

The remainder of this paper is organized as follows: Section \ref{section:related_work} reviews related work in VNE and mobile edge computing. Section \ref{section:methodology} presents our methodology including system model and three allocation strategies. Section \ref{section:experiment} describes our experimental setup and evaluation metrics. Section \ref{section:results} presents results and discussion. Finally, Section \ref{section:conclusion} concludes and outlines future work.

\section{Related Work}
\label{section:related_work}

\begin{figure*}[!t]
\centering
  \includegraphics[width=\linewidth]{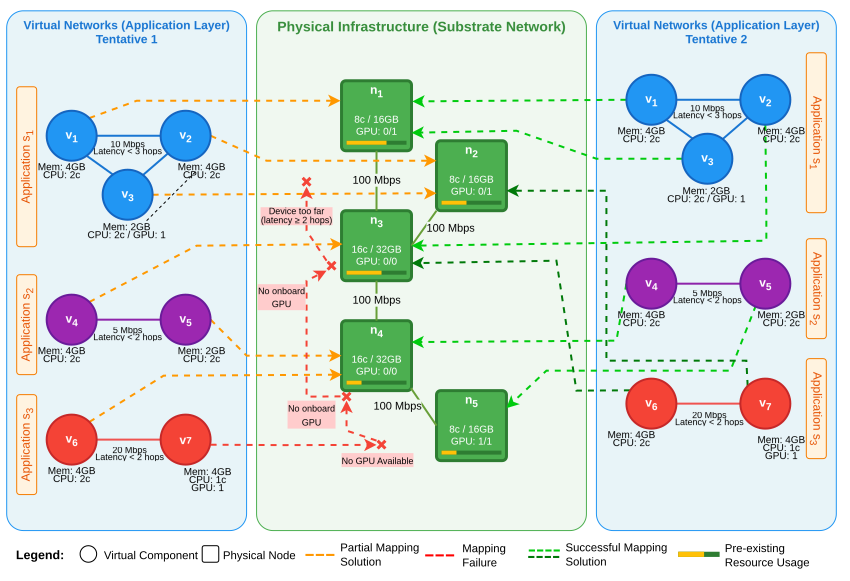}
\caption{Virtual Network Embedding in mobile edge networks. The middle shows virtual network requests (S1, S2, S3) with specific resource demands (CPU, memory, GPU) and bandwidth requirements. The sides depict the physical substrate mesh network of edge devices with limited resources. Arrows indicate partial mappings (orange), failed attempts (red) due to resource constraints (GPU not available on node NS, Latency $\geq$ 2 hops for N1-N4) and successful mappings (green) possible by taking into account the overall system (NSGA-II, ILP) instead of following a greedy approach.}
\label{fig:vne-embedding}
\end{figure*}

This section reviews key areas relevant to extending VNE to highly mobile environments, including traditional VNE approaches, Virtual Network Function (VNF) placement, network slicing, clustering, fault tolerance in decentralized networks, and multi-objective optimization approaches.

Traditional VNE maps virtual networks onto physical substrates, optimizing resources while meeting Quality of Service requirements. Chowdhury and Boutaba \cite{5183468} introduced heuristic algorithms for virtual network mapping, later improving efficiency through coordinated node and link mapping in ViNEYard \cite{5951812}. Fan et al. \cite{9415134} and He et al. \cite{10223368} explored distributed and adaptive VNE strategies for dynamic networks, though mainly in cloud settings with limited mobility. Yu et al. \cite{10.1145/1355734.1355737} and Ebrahimi et al. \cite{9110322} studied VNE in large-scale datacenters, with insights applicable to multi-tiered edge environments.

However, these approaches assume relatively static network topologies and do not address the unique constraints of highly mobile environments, creating a gap that our work addresses.

Similarly, VNF placement optimizes resource utilization in NFV-enabled environments. Bari et al. \cite{7469866} outlined key challenges, while Moens and De Turck \cite{7014205} proposed integer linear programming models for optimal placement. Castellano et al. \cite{8737532} introduced a distributed orchestration algorithm for edge computing, offering scalable solutions relevant to mobile edge deployments.

Multi-objective optimization approaches have gained prominence in resource allocation problems. Genetic algorithms such as Non-dominated Sorting Genetic Algorithm II (NSGA-II), introduced by Deb et al. \cite{996017}, have become cornerstone approaches for solving complex optimization problems with conflicting objectives. Tavakoli-Someh et al. \cite{Tavakoli2019} demonstrated its effectiveness in multi-objective virtual network function placement, optimizing both resource utilization and quality of service metrics simultaneously.

Our work extends these research areas by addressing the specific challenges of VNE in highly mobile environments, providing a novel framework that accounts for the dynamic nature of mobile mesh networks while maintaining the resource optimization goals of traditional VNE approaches.

\section{Methodology}
\label{section:methodology}

This work focuses on centralized resource allocation approaches for mobile edge networks, where a central controller has global visibility of the network state. This approach enables optimal resource allocation decisions and provides a baseline for understanding VNE performance limits in mobile environments. While the target deployment environment consists of distributed edge devices, the centralized optimization insights will inform future distributed implementations.

We address the mobile VNE problem by modeling the infrastructure as a resource-constrained graph and formulating the allocation as a constrained optimization problem. Figure~\ref{fig:vne-embedding} presents a comprehensive illustration of the Virtual Network Embedding problem, demonstrating how multiple applications must be mapped onto resource-constrained physical infrastructure while satisfying bandwidth and latency requirements.

The VNE problem involves mapping virtual components (e.g., V1-V7) from multiple applications onto physical nodes (e.g., N1-N5) while respecting resource capacities and communication requirements. Figure~\ref{fig:vne-embedding} shows three distinct virtual network requests with varying resource demands and their attempted embeddings onto a heterogeneous mesh network. The challenge is particularly acute in mobile environments where devices have limited resources and dynamic availability. Application S3's failure to deploy in the partial mapping case \textemdash due to node N5's resource exhaustion and latency constraints \textemdash exemplifies the critical need for intelligent allocation strategies. This is especially relevant considering that the network could accommodate V7 with a different allocation solution.

To evaluate the effectiveness of different allocation strategies, we employ three distinct methods: greedy heuristics, optimization-based approaches, and multi-objective genetic algorithms. Our approach is validated through simulations across various network topologies, assessing performance under diverse application loads and dynamic network scenarios.

\subsection{System Model}
\label{subsection:model}

\subsubsection{System Parameters}

For this section, we define our set of resource attributes as $\mathcal{A} = \{\text{GPU}, \text{CPU}, \text{Memory}, \text{Storage}\}$. We represent the infrastructure as a substrate network, modeled by the graph $G = (N, E, C)$, where $N$ represents the set of substrate nodes, each corresponding to an embedded device with resource attribute capacities $C_n^a, \forall a \in \mathcal{A}$. The set $E$ contains substrate edges representing communication links between nodes, each with bandwidth capacity $C_e^{BW}$ (in Mbps) and latency $C_e^L$ (in ms).

Our application set is defined as a set of Virtual Resource Requests $(V_s, F_s, L)$ with application $s \in S$. Each virtual node $v \in V_s$ corresponds to an application component that needs to be deployed on a substrate node, with resource demands $L_v^a, \forall a \in \mathcal{A}$. Virtual edges $f \in F_s$ represent communication requirements between virtual nodes, with bandwidth needs $L_f^{BW}$ and latency requirements $L_f^L$.

For simplicity, we define $F = \bigcup_{s \in S} F_s$ and $V = \bigcup_{s \in S} V_s$ as the union of all virtual edges and nodes, respectively. Additionally, we define $F^{+}_v = \{f \in F \mid \exists v \in V, f = (v, *)\}$ and $F^{-}_v = \{f \in F \mid \exists v \in V, f = (*, v)\}$ as the sets of outgoing and incoming edges for virtual node $v$.

\subsubsection{Preprocessing Parameters}

We define $\mathcal{P}_{nm}$ as the set of $k$-shortest paths between node $n$ and node $m$. We denote the union of all such path sets as $\mathcal{P} = \bigcup_{(n, m) \in N^2} \mathcal{P}_{nm}$. For each path $p \in \mathcal{P}$, we denote $o_p = n$ as the origin of path $p$ and $d_p = m$ as the destination of path $p$.
For a given latency requirement $L_f^L$, we will define the set of available paths: $$ \mathcal{P}_f = \{p \in \mathcal{P} \mid \sum_{e \in p} C_e^L \leq L_f^L\} $$

\subsubsection{Variables}

We define decision variables as follows:
\begin{itemize}
    \item $x_s, \forall s \in S$ with $x_s = 1$ if the corresponding application is deployed in the infrastructure
    \item $x_{nv}, \forall n \in N, \forall v \in V$ with $x_{nv} = 1$ if virtual component $v$ is mapped on node $n$
    \item $y_{pf}, \forall p \in \mathcal{P}, \forall f \in F$ with $y_{pf} = 1$ if virtual link $f$ is mapped to physical path $p$
\end{itemize}

\subsubsection{Complete Optimization Formulation}

We formulate the problem as maximizing successful allocations while minimizing normalized latency:

\begin{equation}
	\text{Maximize} \quad \sum_{s \in S} \left( r_s \cdot x_s - \frac{\alpha}{|F_s|} \sum_{f \in F_s} \sum_{p \in \mathcal{P}_f}  y_{pf}  \frac{\sum_{e \in p} C_e^L}{L_f^L}\right)
    \label{eq:objective}
\end{equation}

\textbf{Subject to:}
\begin{align}
\sum_{n \in N} x_{nv} &= x_s, \quad \forall v \in V_s, \forall s \in S \label{eq:integrity_1}\\
\sum_{p \in \mathcal{P}_f} y_{pf} &= x_s, \quad \forall f \in F_s, \forall s \in S \label{eq:integrity_2}\\
\sum_{\substack{p \in \mathcal{P}_f \\ o_p=n}} y_{pf} &= x_{nv}, \quad \forall f \in F^{+}_{v}, \forall n \in N, \forall v \in V \label{eq:integrity_3}\\
\sum_{\substack{p \in \mathcal{P}_f \\ d_p=n}} y_{pf} &= x_{nv}, \quad \forall f \in F^{-}_{v}, \forall n \in N, \forall v \in V \label{eq:integrity_4}\\
\sum_{v \in V} L_v^a \cdot x_{nv} &\leq C_n^a, \quad \forall a \in \mathcal{A}, \forall n \in N \label{eq:node_capacity}\\
\sum_{f \in F} \sum_{\substack{p \in \mathcal{P}_f \\ e \in p}} L_f^{BW} \cdot y_{pf} &\leq C_e^{BW}, \quad \forall e \in E \label{eq:link_capacity}\\
x_s &\in \{0,1\}, \quad \forall s \in S \label{eq:domain_1}\\
x_{nv} &\in \{0,1\}, \quad \forall n \in N, \forall v \in V \label{eq:domain_2}\\
y_{pf} &\in \{0,1\}, \quad \forall p \in \mathcal{P}, \forall f \in F \label{eq:domain_3}
\end{align}

where $r_s$ is a reward score determined as  $r_s = 2 ^{(t-t_s)/\tau}$ with $t$ time when processing, $t_s$ application arrival time and $\alpha \in \mathbb{R}^+$. This ensures that the system will prioritize deploying an application that previously failed over new applications.

Equations (\ref{eq:integrity_1}) to (\ref{eq:integrity_4}) guarantee application deployment integrity: equation (\ref{eq:integrity_1}) enforces complete deployment without redundancy, while equations (\ref{eq:integrity_2}), (\ref{eq:integrity_3}), and (\ref{eq:integrity_4}) handle the mapping of virtual links between application components onto the physical infrastructure. 

Node capacity constraints (\ref{eq:node_capacity}) prevent resource overallocation by ensuring that the cumulative demands of all virtual components assigned to each physical device do not exceed its available CPU, memory, GPU, and storage capacities. Link capacity constraints (\ref{eq:link_capacity}) maintain network feasibility by limiting the total bandwidth demands routed through each physical connection, preventing network congestion and ensuring quality of service for deployed applications. Latency requirements are enforced implicitly through path preprocessing, where only paths satisfying end-to-end latency bounds are included in $\mathcal{P}_f$, effectively constraining the solution space to latency-feasible allocations. Constraints (\ref{eq:domain_1})-(\ref{eq:domain_3}) specify the binary domains for all decision variables.

This overall system ensures consistent application deployment. This formulation captures the essential trade-offs between resource utilization efficiency and application performance requirements.

\subsection{Implementation}

For our tests, we experimented with a centralized approach which assumes global network visibility with no information transmission delays between devices, though we simulate deployment latency. This enables optimal solutions using complete system state knowledge.

We implement three allocation strategies: Integer Linear Programming provides theoretical upper bounds and optimal allocation; greedy heuristics offer immediate deployment with lower computational cost; NSGA-II balances multiple objectives through evolutionary optimization.

The ILP serves as a benchmark for system performance limits. The greedy algorithm prioritizes computational efficiency over optimality. NSGA-II simultaneously optimizes conflicting objectives, providing efficient trade-offs between latency and application acceptance.

\section{Experiment}
\label{section:experiment}

We evaluate our three allocation strategies using a custom Python Discrete Event Simulator with NetworkX. The simulator models heterogeneous device networks and processes application requests arriving via Poisson process (rate $\lambda$) with exponential lifetimes (rate $\mu$, where $\mu$ < $\lambda$).

%% Reduced version
All strategies use k-shortest path preprocessing filtered by latency requirements. Upon application departure, all allocated resources are released. Failed deployments retry after delay $\tau$ with increasing priority until maximum retries are reached.

%% Longer version
%For both allocation strategies, we use path preprocessing to reduce computational complexity. The system pre-computes k-shortest paths between all node pairs using NetworkX's k-shortest path algorithm. These paths are filtered based on latency requirements, creating a viable path set for each potential virtual link.

All three allocation strategies process incoming application requests, which arrive according to a Poisson process with rate $\lambda$. Each request consists of multiple application components with specific resource requirements and inter-component communication needs. The allocation algorithms determine a placement for these components across the network, considering resource constraints and minimizing component-to-component latency.

Once deployed, applications remain active until the end of their lifetime, following an exponential distribution with rate $\mu$ (where $\mu < \lambda$ to gradually increase system load). 

Upon departure, all resources allocated to the application are released. All algorithms operate on identical random scenarios defined by setting a common seed, ensuring fair comparison.

The discrete event simulator handles the event sequence chronologically, simulating deployment delays, resource allocation, and back-off mechanisms for failed deployments. If a placement fails due to insufficient resources, the system attempts redeployment after a specific delay. After a successive failures (Table \ref{table:simulation}), the application is considered rejected.

\subsection{Reference Allocation}
\label{subsec:reference}

The greedy approach based on Yu et al. \cite{10.1145/1355734.1355737} processes applications in batches with the following allocation strategy: applications are sorted by arrival time (oldest first), components are ranked by resource requirements (highest first), and devices are ranked by proximity to origin, resource availability, and current load. Components are placed on the highest-ranked suitable devices, followed by bandwidth allocation for inter-component links using shortest available paths. Failed placements trigger retry mechanisms with exponential back-off up to a maximum retry limit, after which applications are rejected.

This greedy approach offers low computational complexity but may lead to suboptimal solutions in terms of both application placement and latency minimization compared to global optimization approaches.
\subsection{Optimization-Based Allocation}

We implement our ILP based allocation strategy to determine optimal resource allocation under the same experimental conditions. Unlike the greedy approach, this method considers the global system state to find the optimal solution for each batch of requests.

The ILP formulation from Section \ref{section:methodology} is solved using Gurobi (v12.0.1) with time limit $\tau$. Successful allocations trigger deployment events; failed applications retry with increasing priority. This approach should provide a more efficient solution compared to the reference allocation at the cost of higher computational complexity.

\subsection{Multi-Objective Genetic Algorithm (NSGA-II)}
\label{subsec:nsga2}

NSGA-II implementation \cite{996017} balances application acceptance and latency minimization. Parameters: 50 individuals, 100 generations, 0.8 crossover rate, 0.1 mutation rate.

\subsection{Performance Metrics}

To evaluate all three allocation strategies, we measure application deployment metrics, resource utilization and network performances.  Our application deployment metrics include acceptance ratio as percentage of successfully deployed applications relative to total requests, as well as rejection rate after multiple retries, and statistics regarding the batch sizes. 

Resource utilization is tracked and stored at all times for all devices down to millisecond precision, allowing us to plot per-device utilization over time, statistical distributions, and to monitor average utilization across all devices to monitor resource balance across the network. 

Network performances are measured through average bandwidth utilization on physical links as well as normalized latency for deployed applications, both expressed in percentage of their maximal theoretical value.

These metrics are captured in real-time in the simulator, allowing for both temporal analysis and aggregated comparison. This metrics collection enables detailed comparison between the allocation strategies under varying network conditions and application workloads. It also facilitates the identification of specific scenarios where each approach excels, and provides insights for potential hybrid approaches in future work.

\subsection{Experimental Parameters}
\label{subsubsection:scenarios}

We simulated a network of 20 devices randomly placed in a 3D space linked with simulated wireless bridges. All topologies satisfy a minimum connectivity requirement ensuring that every device can reach every other device through multi-hop routing. Links between devices, when any, are 100 MB/s links.

%% Reduced size
The substrate network consists of two device types with heterogeneous resource capacities: a 16-core CPU device with 32 GB of RAM, or an 8-core CPU device with 16 GB of RAM and a GPU with 8 GB of VRAM. Both device types include 1 TB of storage.
%% Longer version
%The substrate network consists of two types of devices with heterogeneous resource capacities detailed in Table \ref{table:device}.

%\begin{table}[h]
%\centering
%\caption{Device Configuration Parameters}
%\label{table:device}
%\begin{tabular}{|l|c|c|}
%\hline
%\textbf{Resource} & \textbf{Type 1 Device} & \textbf{Type 2 Device} \\
%\hline
%CPU & 16 cores & 8 cores \\
%\hline
%Memory & 32 GB & 16 GB \\
%\hline
%Storage & 1 TB & 1 TB \\
%\hline
%GPU & 0 & 8 units \\
%\hline
%\end{tabular}
%\end{table}

Applications consist of multiple interconnected components with specific resource requirements detailed in Table \ref{table:application}.

\begin{table}[h]
\centering
\caption{Application Characteristics}
\label{table:application}
\begin{tabular}{|l|l|}
\hline
\textbf{Parameter} & \textbf{Distribution} \\
\hline
Components per application & 1-5 (uniform) \\
\hline
CPU request per component & 0.1-2 cores (uniform) \\
\hline
Memory request per component & 100-4096 MB (uniform) \\
\hline
Storage request per component & 1-250 GB (uniform) \\
\hline
GPU request per component & 0-2 units (50\% zero) \\
\hline
Bandwidth request per link & 5-20 MB/s (uni.)\\
\hline
\end{tabular}
\end{table}

Component-to-component links are established using a probabilistic approach that ensures connectivity while modeling realistic application topologies. Each component is first linked to its predecessor in the application, ensuring a connected graph structure. Additionally, each component may establish connections to other components with a probability of $p = 1/n$, where $n$ is the total number of components in the application, creating a mix of linear, star, and mesh topologies that reflect real-world application architectures while maintaining reasonable bandwidth requirements across the network.

\begin{table}[h!]
\centering
\caption{Simulation Parameters}
\label{table:simulation}
\begin{tabular}{|l|l|}
\hline
\textbf{Parameter} & \textbf{Value} \\
\hline
Arrival rate ($\lambda$) & 120 applications/hour \\
\hline
Departure rate ($\mu$) & 45 applications/hour \\
\hline
Simulation duration & 4 hours \\
\hline
Retry timeout & 16 minutes \\
\hline
Maximum retries & 15 \\
\hline
Batch interval ($\tau$) & 1 minutes \\
\hline
Application reward function & $2 ^{(t-t_s)/\tau}$ \\
\hline
Balancing parameter ($\alpha$)& $2^{10}$ \\
\hline
\end{tabular}
\end{table}

These parameters were selected to represent realistic workloads in edge computing environments while enabling stress-testing of the allocation algorithms. The arrival and departure rates create a gradually increasing system load, allowing us to observe behavior under varying resource pressure across all three approaches.

\section{Results and Discussion}
\label{section:results}

This section presents the results of our experiments and analyzes their implications for application deployment in mesh edge networks. We examine three key aspects of system performance: latency optimization, application deployment metrics, and resource utilization consumption patterns across all three allocation strategies: Greedy, NSGA-II, and ILP.

In most of our analysis, we chose to focus primarily on the steady-state behavior of the system by omitting the initial transient period (the first 60 minutes), but we are still presenting the transient period as part of our plots for additional information. This approach allows us to evaluate the long-term performance characteristics of the allocation strategies under load, which is more representative of real-world operational conditions. During the transient period, the system experiences initialization effects and has not yet reached typical operating conditions, potentially skewing the comparative analysis.

The results are smoothed over a 5-minute window to balance spikes, and the shaded area on both sides of the curves represent 10-90 percentile confidence intervals.

\subsection{Application Deployment Performance}
\label{subsection:acceptance}

\begin{figure}[h!]
    \centering
   \includegraphics[width=\linewidth]{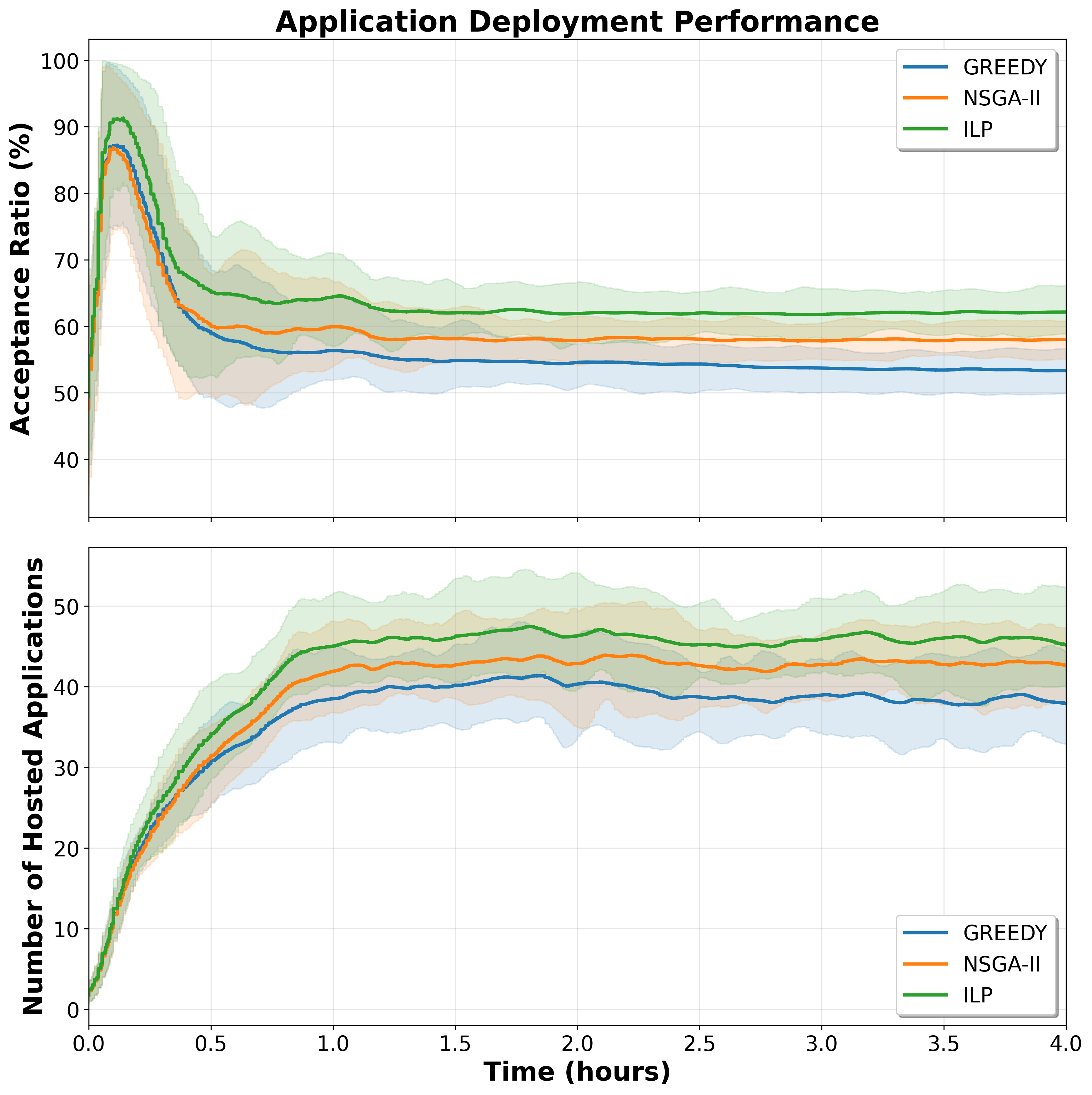}
    \caption{Performance comparison of three allocation strategies showing (top) acceptance ratios over time, and (bottom) concurrent application counts}
    \label{fig:acceptance}
\end{figure}

Figure~\ref{fig:acceptance} shows the evolution of the acceptance ratio and the number of concurrent hosted applications over time. The results follow similar evolution between our three approaches.

The ILP approach consistently achieves higher acceptance ratios throughout the simulation period, maintaining approximately 62\% acceptance while the reference allocation shows acceptance ratio around 53\%. NSGA-II achieves competitive performance with around 58\% acceptance.

In terms of concurrent hosted application. The ILP achieves the highest number, maintaining around 45 application, followed closely by the NSGA-II with around 43 applications. The simpler reference allocation stabilizes at approximately 39 applications.

The superior performance from both the ILP and NSGA-II can be attributed to their global optimization perspective, which enable better resource utilization across the entire network, particularly in early stages, and to prioritize applications effectively with regards to their expiration delay after multiple failed attempts.

\subsection{Latency Performance}
\label{subsection:latency}

\begin{figure}[h!]
  \centering
  \includegraphics[width=\linewidth]{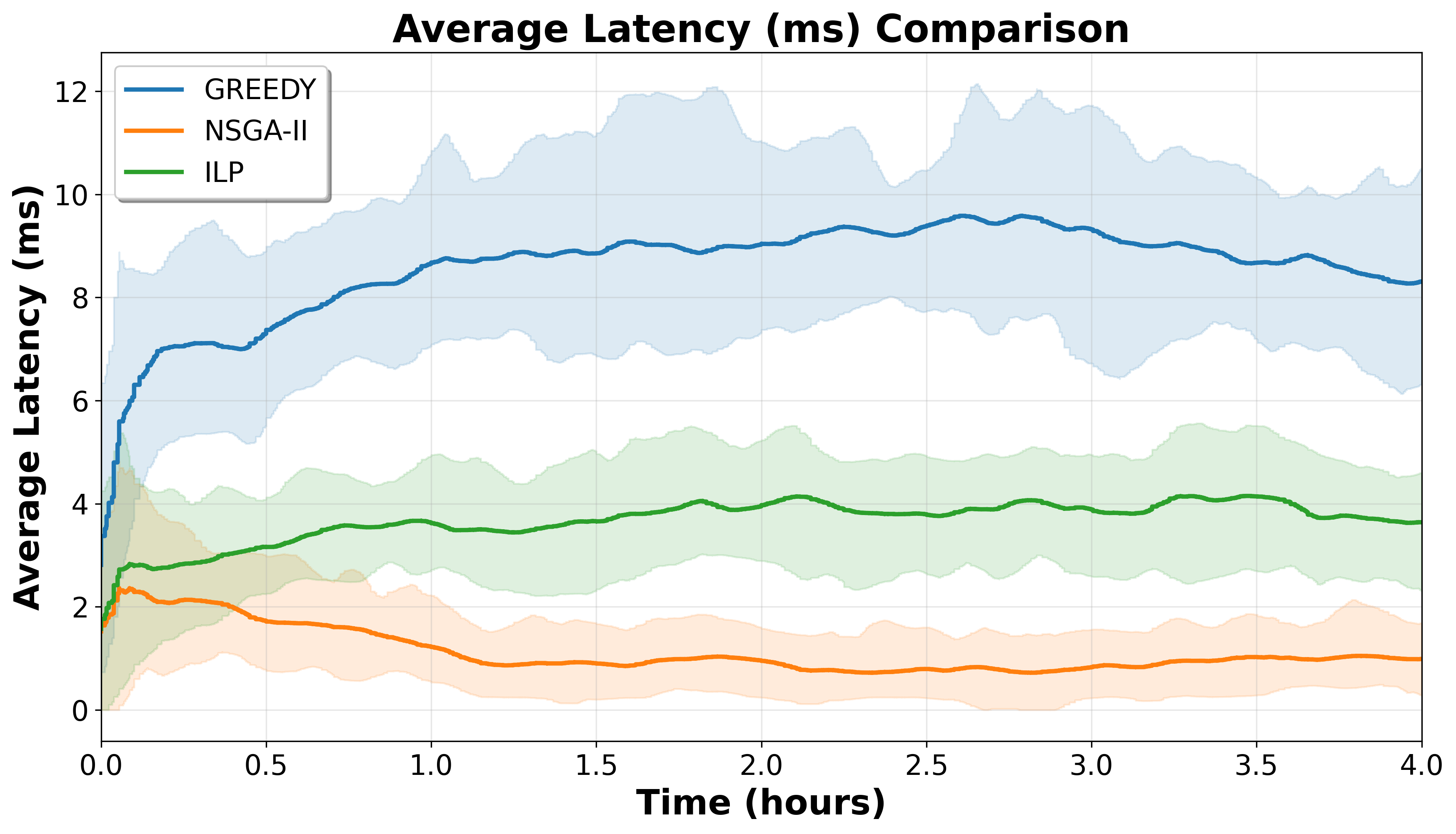}
  \caption{Average latency comparison across allocation strategies}
  \label{fig:latency_comparison}
\end{figure}

Figure~\ref{fig:latency_comparison} demonstrates significant differences in terms of latency optimization capabilities among the three algorithms. The results show a clear distinction between algorithms.

NSGA-II achieves the best latency performance with an average of 1ms in steady state, which comes from its capacity to dynamically adjust between acceptance ratio and latency minimization compared to the ILP which prioritizes acceptance through the choice of a higher reward in the application acceptance. Changing the $\alpha$ parameter in the ILP can change this balance between application acceptance and latency with higher values resulting in similar results, lowering both latency and acceptance ratio.

Latency is computed considering that 1-hop between two devices amounts to around 10ms and that two application components on the same device incur no latency. The value is then averaged on all links within each application then averaged on the number of applications in the system to obtain an average latency per link between components per application.

The ILP shows moderate latency improvement, cutting down the average latency by approximately 50\% compared to the reference allocation, averaging around 4ms. This demonstrates the optimization approach's capability to either group application components on a single device or provide placements that require fewer hops between application components. Overall, the optimization-based approach effectively minimizes communication distances between interdependent application components compared to the reference allocation.

The greedy reference allocation exhibits the highest latency at approximately 8.5ms, reflecting its local decision-making approach that prioritizes immediate placement feasibility over global communication optimization.

Lower latency has practical implications for mesh network deployments as it minimizes end-to-end application latency, which is crucial for time-sensitive applications. Additionally, reduced communication distances in wireless mesh networks should translate to lower power consumption for wireless communication, extending the operational lifetime of battery-powered mesh devices, which we could verify as further work on Humanitas's systems.

This highlights how intelligent placement strategies can significantly impact key network parameters such as latency while maintaining similar overall resource utilization patterns.

\subsection{Resource utilization}
\label{subsection:comprehensive_view}

\begin{figure}[h!]
  \centering
  \includegraphics[width=\linewidth]{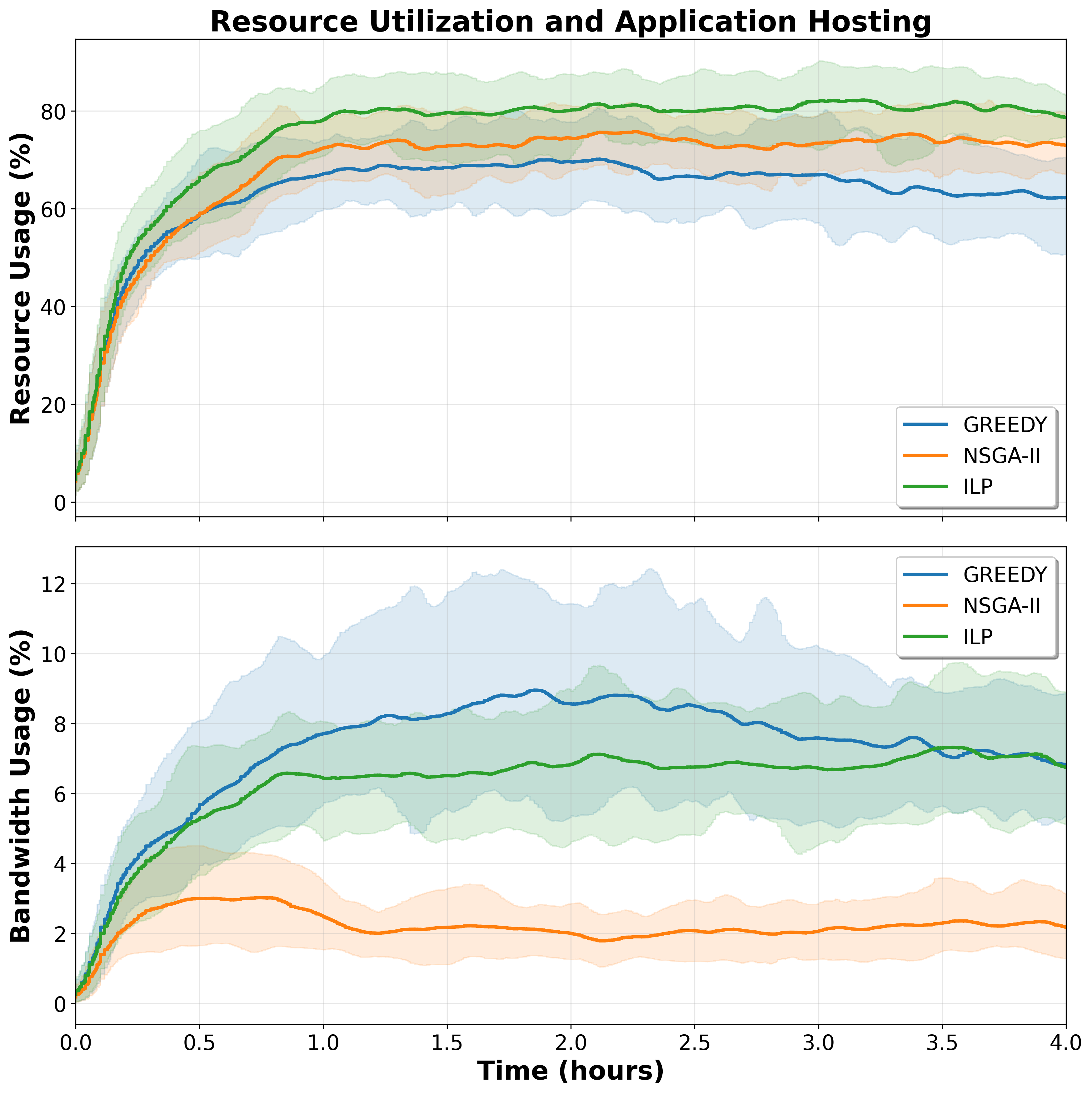}
  \caption{Resource utilization patterns showing (top) bandwidth usage efficiency, and (bottom) maximum device resource consumption}
  \label{fig:output_avg_resource}
\end{figure}

Figure~\ref{fig:output_avg_resource} presents bandwidth usage patterns across the three algorithms. The bandwidth utilization shows similar variations compared to latency.

The greedy reference algorithm shows the highest bandwidth utilization with a steady state around 8.5\%, the algorithm uses longer, less optimal, paths that consume more network resources. The ILP maintains moderate bandwidth usage, reflecting its balanced approach. While NSGA-II demonstrates the most efficient bandwidth utilization at around 2.5\% of the overall system, showing that NSGA-II collocates application components more that the other two.

Maximum resource usage shows expected results with regards to application allocation, with the system hosting the most concurrent apps uses the most resources. 

To dig further into the numbers, the transient period, within the first hour, shows similar results initially before diverging, with the ILP and NSGA-II allocations allocating more resources. Once this 1-hour mark is reached, the system arrives at steady state where the ILP maintains a 80\% maximum resource use, the NSGA-II reaches 73\% and the Reference allocation stabilizes around  65\% maximum resource use.

This difference can be explained by the variation in the number of concurrent application running in the system, which, albeit small (a 5\% difference between each method and the next in their rankings), should still amount for differences in resource use. Another explanation lies in application deployment choices. The ILP approach places different applications compared to the reference allocation, which accounts for another part of the difference in resource utilization. The beginning of the transient period, with very few applications hosted and almost identical resource use, confirms that hosting identical applications results in similar patterns.

\subsection{Implications for Mesh Edge Networks}
\label{subsection:additional_discussion}

These results have several implications for the deployment of applications in mesh edge networks, particularly in scenarios where infrastructure reliability and optimal resource utilization are critical.

The choice of algorithm should depend on deployment priorities. The ILP is optimal when maximizing application acceptance is paramount, NSGA-II excels when balancing multiple objectives more dynamically including latency.

While optimization-based approaches require more computational resources, the significant performance improvements justify this cost for non-urgent deployments. The 10-50\% improvements in key metrics can be critical in emergency scenarios, especially considering that networking is not impacted much, however this would require further anaylisis on appplication network utilization patterns to confirm our theory.

All three algorithms demonstrate stable performance under increasing load, suggesting they would maintain reliability during extended emergency operations

While our optimization approach shows clear advantages, it does require more computational resources for solving the ILP. In practical deployments, this trade-off must be considered against the improved application deployment efficiency. The optimization process is typically run in the Core Network to ensure minimal resource consumption for the limited worker nodes, which might prove troublesome in edge-only infrastructures. For highly resource-constrained environments, a hybrid approach might be optimal, exploiting the flexibility behind the NSGA-II allocation when the trade-off between resources and latency is sensitive and the optimization-based approach for planned application placements or for migrations.

\section{Conclusion and Future Work}
\label{section:conclusion}

We evaluated three resource allocation strategies for constrained mesh edge networks through comprehensive simulation. Our results demonstrate distinct advantages: ILP provides optimal application acceptance, NSGA-II excels in multi-objective balancing with flexible latency performance, and greedy approaches offers computational efficiency for immediate deployment scenarios. Our optimization framework establishes theoretical performance bounds for future distributed implementations.

Performance differences range from 10x latency improvements to 15\% acceptance gains, highlighting algorithm selection importance in constrained environments. While optimization approaches require more computation, substantial performance improvements justify this cost for critical deployments.

Future work will distribute allocation across mesh networks for autonomous edge operations, address mobility and energy consumption, and validate solutions on real hardware deployments. Machine learning approaches are also to be explore to offer additional perspectives on resource allocation in constrained environments.

This research provides valuable insights for developing robust, autonomous systems capable of efficient operation when traditional infrastructure is compromised.

%% Longer version
\section{Acknowledgments}
This research was supported by MITACS through the MITACS Accelerate program, under project IT30130.

\bibliographystyle{unsrt}
\bibliography{biblio}

\end{document}